\documentclass[conference]{IEEEtran}
\IEEEoverridecommandlockouts
\usepackage{cite}
\usepackage{amsmath,amssymb,amsfonts}
\usepackage{algorithmic}
\usepackage{graphicx}
\usepackage{textcomp}
\usepackage{xcolor}
\usepackage{hyperref} 
\usepackage{setspace}

\def\BibTeX{{\rm B\kern-.05em{\sc i\kern-.025em b}\kern-.08em
    T\kern-.1667em\lower.7ex\hbox{E}\kern-.125emX}}
    
\usepackage{enumitem}
\usepackage{bm}

\ifCLASSOPTIONcompsoc
\usepackage[caption=false,font=normalsize,labelfont=sf,textfont=sf]{subfig}
\else
\usepackage[caption=false,font=footnotesize]{subfig}
\fi

\usepackage{multirow}
\usepackage{tabularx,stackengine,collcell}
\let\endminwd\relax
\newcolumntype{L}[1]{>{\collectcell\xminwd l{#1}}l<{\endminwd\endcollectcell}}
\newcolumntype{C}[1]{>{\collectcell\xminwd c{#1}}c<{\endminwd\endcollectcell}}
\newcolumntype{R}[1]{>{\collectcell\xminwd r{#1}}r<{\endminwd\endcollectcell}}
\def\minwd#1#2#3\endminwd{\stackengine{0pt}{#3}{\rule{#2}{0pt}}{O}{#1}{F}{F}{L}}
\newcommand\xminwd[1]{\minwd#1}

\newcolumntype{Q}{@{}c@{}}
\newcommand{\smallsection}[1]{{\vspace{0.02in} \noindent {\bf{\underline{\smash{#1}.}}}}}
\newcommand{\kijung}[1]{{\color{black}#1}}
\newcommand{\noseong}[1]{{\color{black}#1}}
\newcommand{\jihoon}[1]{{\color{black}#1}}

\newcommand{\GG}{\mathcal{G}}

\newcommand{\V}{\mathcal{V}}
\newcommand{\E}{\mathcal{E}}

\newtheorem{definition}{Definition}
\newtheorem{theorem}{Theorem}
\newtheorem{proposition}{Proposition}
\usepackage{flushend}

\usepackage{fancyhdr}
\usepackage{kantlipsum}
\fancypagestyle{plain}{
\fancyhf{}
\fancyhead[C]{Conference on \LaTeX} 

}
\usepackage{eso-pic}
\begin{document}
\AddToShipoutPictureBG*{
\AtPageUpperLeft{
\setlength\unitlength{1in}
\hspace*{\dimexpr0.5\paperwidth\relax}
\makebox(0,-0.75)[c]{\textbf{2020 IEEE/ACM International Conference on Advances in Social
Networks Analysis and Mining (ASONAM)}}}}


\title{MONSTOR: An Inductive Approach for Estimating and Maximizing Influence over Unseen Networks}

\author{\IEEEauthorblockN{Jihoon Ko}
\IEEEauthorblockA{KAIST AI \\
jihoonko@kaist.ac.kr}
\and
\IEEEauthorblockN{Kyuhan Lee}
\IEEEauthorblockA{KAIST AI \\
kyuhan.lee@kaist.ac.kr}
\and
\IEEEauthorblockN{Kijung Shin}
\IEEEauthorblockA{KAIST AI \& EE \\
kijungs@kaist.ac.kr}
\and
\IEEEauthorblockN{Noseong Park}
\IEEEauthorblockA{Department of AI, Yonsei University \\
noseong@yonsei.ac.kr}
}
\maketitle


\begin{abstract}
Influence maximization (IM) is one of the most important problems in social network analysis. Its objective is to find a given number of seed nodes that maximize the spread of information through a social network. Since it is an NP-hard problem, many approximate/heuristic methods have been developed, 
and a number of them repeat Monte Carlo (MC) simulations over and over to reliably estimate the influence (i.e., the number of infected nodes) of a seed set.
In this work, we present an inductive machine learning method, called \underline{Mon}te Carlo \underline{S}imula\underline{tor} (MONSTOR), for estimating the influence of given seed nodes in social networks unseen during training. 
To the best of our knowledge, MONSTOR is the first inductive method for this purpose.
MONSTOR can greatly accelerate existing IM algorithms by replacing repeated MC simulations. 
In our experiments, MONSTOR provided highly accurate estimates, achieving \jihoon{0.998} or higher Pearson and Spearman correlation coefficients in unseen real-world social networks.
Moreover, IM algorithms equipped with MONSTOR are more accurate than state-of-the-art competitors in \jihoon{63\%} of IM use cases.
\end{abstract}


\section{Introduction}

{\it Influence maximization} (IM) \cite{Kempe:2003:MSI:956750.956769} is to find a certain number of seed nodes who maximize the spread of information through a social network. It is NP-hard, and many IM algorithms (e.g., Greedy~\cite{Kempe:2003:MSI:956750.956769}, CELF~\cite{Goyal:2011:COG:1963192.1963217}, and UBLF~\cite{6729575}) rely on repeated Monte Carlo (MC) simulations of information cascade processes. An MC simulation takes $\mathcal{O}(|\E|)$ time, where $|\E|$ is the number of edges, and estimating the influence of a seed set via $d$ simulations takes $\mathcal{O}(d|\E|)$ time, which is one of the main performance bottlenecks. In~\cite{Kempe:2003:MSI:956750.956769,6729575}, $d$ is set to $10,000$.

In this work, we propose a neural network-based method, called \underline{Mon}te Carlo \underline{S}imula\underline{tor} (MONSTOR), for estimating MC simulation results under the independent cascade (IC) model. MONSTOR is inductive. That is, it is capable of estimating MC simulation results in social networks unseen during training. To the best of our knowledge, no existing IM method is inductive. That is, existing methods cannot be used for estimating influences in unseen social networks~\cite{DBLP:conf/atal/YanSLMLS19,DBLP:journals/corr/abs-1906-07378}. After being trained, MONSTOR significantly speeds up existing IM methods by replacing repeated MC simulations. 

\begin{figure}[t]
    \vspace{-4mm}
    \centering
    \subfloat[][]{\includegraphics[width=0.35\columnwidth]{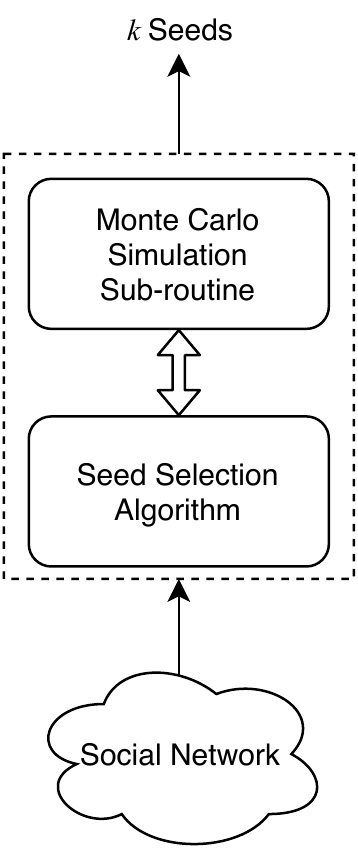}}
    \subfloat[][]{\includegraphics[width=0.35\columnwidth]{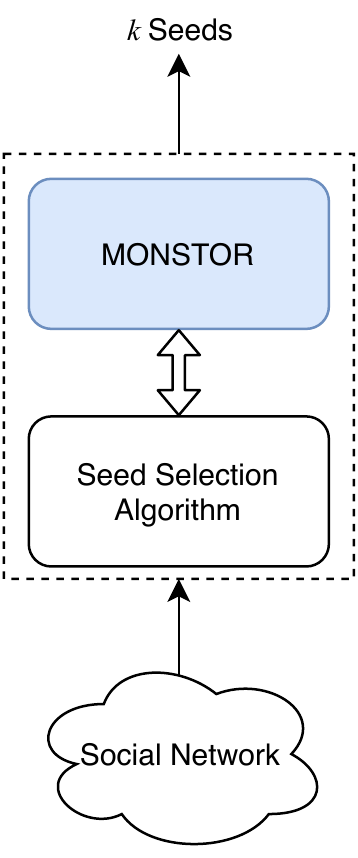}}
    \caption{Comparison of Influence Maximization Approaches: (a) MC simulated-based approaches, and (b) our approach equipped with the proposed \underline{Mon}te Carlo \underline{S}imula\underline{tor} (MONSTOR).}
    \label{fig:archi}
\end{figure}

We conduct experiments with three real-world social networks. One strong point in our experiments is that we use real activation probabilities of edges, which are calculated from retweet logs. That is, we weight each directed edge $(u,v)$ with the real probability that user $u$ influences $v$. Note that most previous studies on influence maximization simply used random, uniform, and degree-based
probabilities~\cite{10.1145/2882903.2915207,DBLP:journals/corr/abs-1111-4795,pmia,Goyal:2011:COG:1963192.1963217,6729575}, which are different from real ones.

In our experiments, MONSTOR yielded near-perfect estimations in terms of the Pearson's correlation coefficients and the Spearman's Rank correlation coefficients. In addition, simulation-based IM algorithms \cite{Kempe:2003:MSI:956750.956769,Goyal:2011:COG:1963192.1963217,6729575} equipped with MONSTOR yielded almost the same influence maximization results as those of the original algorithms based on MC simulations.
Moreover, they were more accurate than state-of-the-art non-simulation-based IM algorithms \cite{10.1145/2882903.2915207,DBLP:journals/corr/abs-1111-4795,pmia} in 17 out of the 27 cases in our experiments. 

The rest of this paper is organized as follows. In Section \ref{sec:concepts}, we introduce some concepts and notions related to our problem. In Section \ref{sec:method}, we describe the overall workflow and the detailed design of MONSTOR. In Section \ref{sec:exp}, we present experimental results. After reviewing previous related work in Section~\ref{sec:rel}, we briefly discuss the application of our approach to the linear threshold model in Section~\ref{sec:lt}. We offer conclusions in Section~\ref{sec:conclusion}.

\section{Concepts and Problem Definition}
\label{sec:concepts}

In this work, we focus on the \textit{independent cascade} (IC) model \cite{Kempe:2003:MSI:956750.956769}, where an infected node $u$ attempts \textbf{once} to activate (i.e., directly infect) each neighbor $v$ and the probability of success is $p_{(u,v)}$. 
This process is repeated for each newly infected node until there are no newly infected nodes.

\begin{definition}[Activation Probability]
\label{defn:prob} The {\it activation probability} $p_{(u,v)}$ from $u$ to $v$ is the success probability that the node $u$ activates 
its neighbor $v$ when $u$ is infected.
\end{definition}

The adjacency matrix when weighting each directional edge $(u,v)$ by $p_{(u,v)}$ is called the {\it activation probability matrix} $\mathbf{P}\in [0,1]^{|\V| \times |\V|}$. That is, each $(u,v)$-th entry \noseong{of} $\mathbf{P}$ is $p_{(u,v)}$.


We use interaction logs such as retweets among users to construct a social network $\GG=(\V,\E)$, and we measure activation probabilities as follows \cite{Goyal:2010:LIP:1718487.1718518,Kempe:2003:MSI:956750.956769}:
\begin{enumerate}[leftmargin=*]
    \item Bernoulli Trial (BT): $$p_{(u,v)} = \frac{|actions(u, *) \cap actions(v, *)|}{|actions(u, *)|},$$
    \item Jaccard Index (JI): $$p_{(u,v)} = \frac{|actions(u, *) \cap actions(*, v)|}{|actions(u, *) \cup actions(*, v)|},$$ 
    \item Linear Probability (LP): $$p_{(u,v)} = \frac{|actions(u, *) \cap actions(*, v)|}{|actions(*, v)|},$$
\end{enumerate} where $actions(x, *)$ denotes the set of actions (e.g., retweets and replies) done by node $x$, and $actions(*, x)$ denotes the set of actions whose object (e.g., author of retweeted tweets and recipient of replies) is $x$.
We consider all these probabilities, and thus we define three different activation probability matrices, $\mathbf{P}_{BT}$, $\mathbf{P}_{JI}$, and $\mathbf{P}_{LP}$, from a social network. 

Based on $\mathbf{P}$, we define the infection probability of each node given a set of \textit{seed nodes} (i.e., initially infected nodes).

\begin{definition}[Infection Probability] \label{defn:infection}
Given a seed set $S\subseteq \V$, for each node $x\in \V$,
the \textit{infection probability} $\rho(x)$ is the probability that $x$ is infected under the IC model with $S$.
\end{definition}

In this work, we consider the following two problems:
\begin{itemize}[leftmargin=*]
    \item \textbf{Influence Estimation (IE)}: 
    \begin{itemize}
    \item \textbf{Given} a seed set $S$, 
    \item \textbf{Estimate} its \textit{influence} $\sum_{x\in \V}\rho(x)$ (i.e., the expected number of infected nodes). 
    \end{itemize}
    \item \textbf{Influence Maximization (IM)} \cite{Kempe:2003:MSI:956750.956769}: 
    \begin{itemize}
    \item \textbf{Given} the number of seed nodes $k$, 
    \item \textbf{Find} the set $S$ of $k$ seed nodes
    \item to \textbf{Maximize}  the influence $\sum_{x\in \V}\rho(x)$.
    \end{itemize}
\end{itemize}

\section{Proposed Method}
\label{sec:method}


In this section, we propose a general inductive model MONSTOR for estimating the infection probability $\rho(x)$ of every node $x\in \V$ in a network that is not necessarily a part of training data (i.e., in a network unseen during training).
Note that we can answer the IE problem by summing the estimates and answer the IM problem by replacing MC simulations in simulation-based algorithms (e.g., \cite{Kempe:2003:MSI:956750.956769,Goyal:2011:COG:1963192.1963217,6729575}) with MONSTOR.

\begin{figure}[t]
    \centering
    \includegraphics[width=\columnwidth]{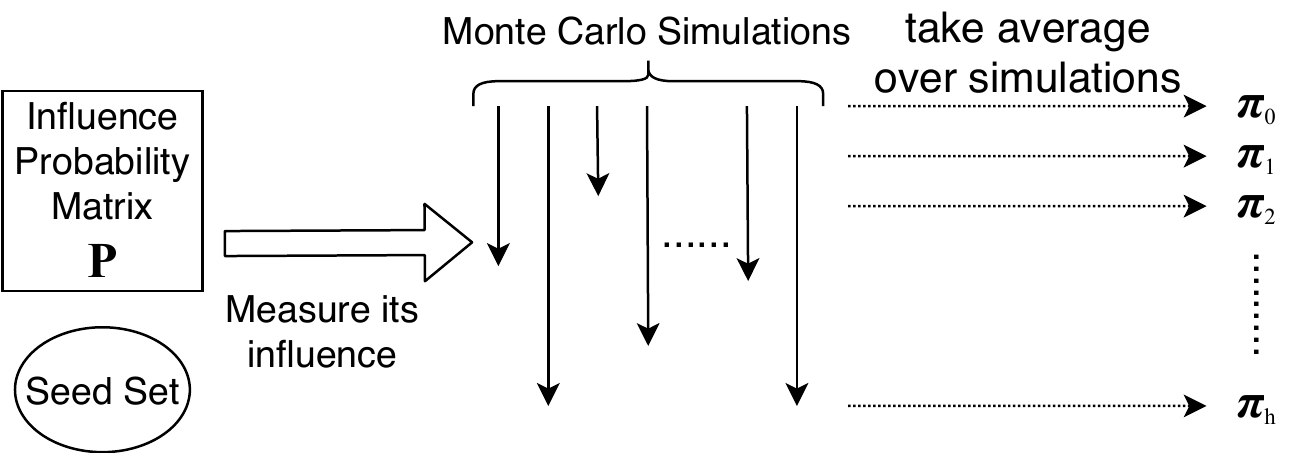}
    \caption{How to build training data. Given a network $\GG$ (whose activation probability matrix is $\mathbf{P}$) and a seed set, we perform multiple simulations and collect $\boldsymbol{\pi}_0$, $\cdots$, $\boldsymbol{\pi}_h$, where $\boldsymbol{\pi}_h=\boldsymbol{\pi}_{h+1}$, for the seed set. Note that the inner product $\langle\mathbf{1},\boldsymbol{\pi}_h\rangle$ is the influence, i.e., the number of infected nodes under the IC model.
    We repeat these steps with many different seed sets. 
    \label{fig:sims}}
\end{figure}

Note that seed nodes are infected in the $0$-th step of the IC model, and those infected directly by seed nodes are infected in the $1$-st step. That is, a step is one-hop cascade  from newly infected nodes.
From Definitions~\ref{defn:limited}-\ref{defn:limited:vector}, Proposition~\ref{pro:mono} follows.

\begin{definition}[Infection Probability within Limited Steps] \label{defn:limited}
Given a seed set $S \subseteq \V$, $\rho_{i}(x)$ denotes the infection probability during the first $i$ steps of the IC model.
Note that $\rho_{i}(x)\approx \rho(x)$ if $i$ is sufficiently large, and $\rho_{i}(x) = \rho(x)$ if $i$ is greater than or equal to the longest path length in the input network.
\end{definition}

\begin{definition}[Infection Probability Vector] \label{defn:limited:vector}
Let $\boldsymbol{\pi}:= [\rho(x)] \in [0,1]^{|\V|}$ be the vector of $\rho(x)$, $\forall x \in \V$. We also let $\boldsymbol{\pi}_i := [\rho_{i}(x)]$. 
\end{definition}

\begin{proposition}
\label{pro:mono}
The infection probability monotonically increases w.r.t. $i$. That is, $\boldsymbol{\pi}_i \leq \boldsymbol{\pi}_{i+1}$, or equivalently, $\rho_{i}(x)\leq\rho_{i+1}(x)$ for all $i\geq 0$ and $x\in \V$.
\end{proposition}

\subsection{Overall Workflow}\label{sec:over}
\begin{figure}[t]
    \centering
    \includegraphics[width=\columnwidth]{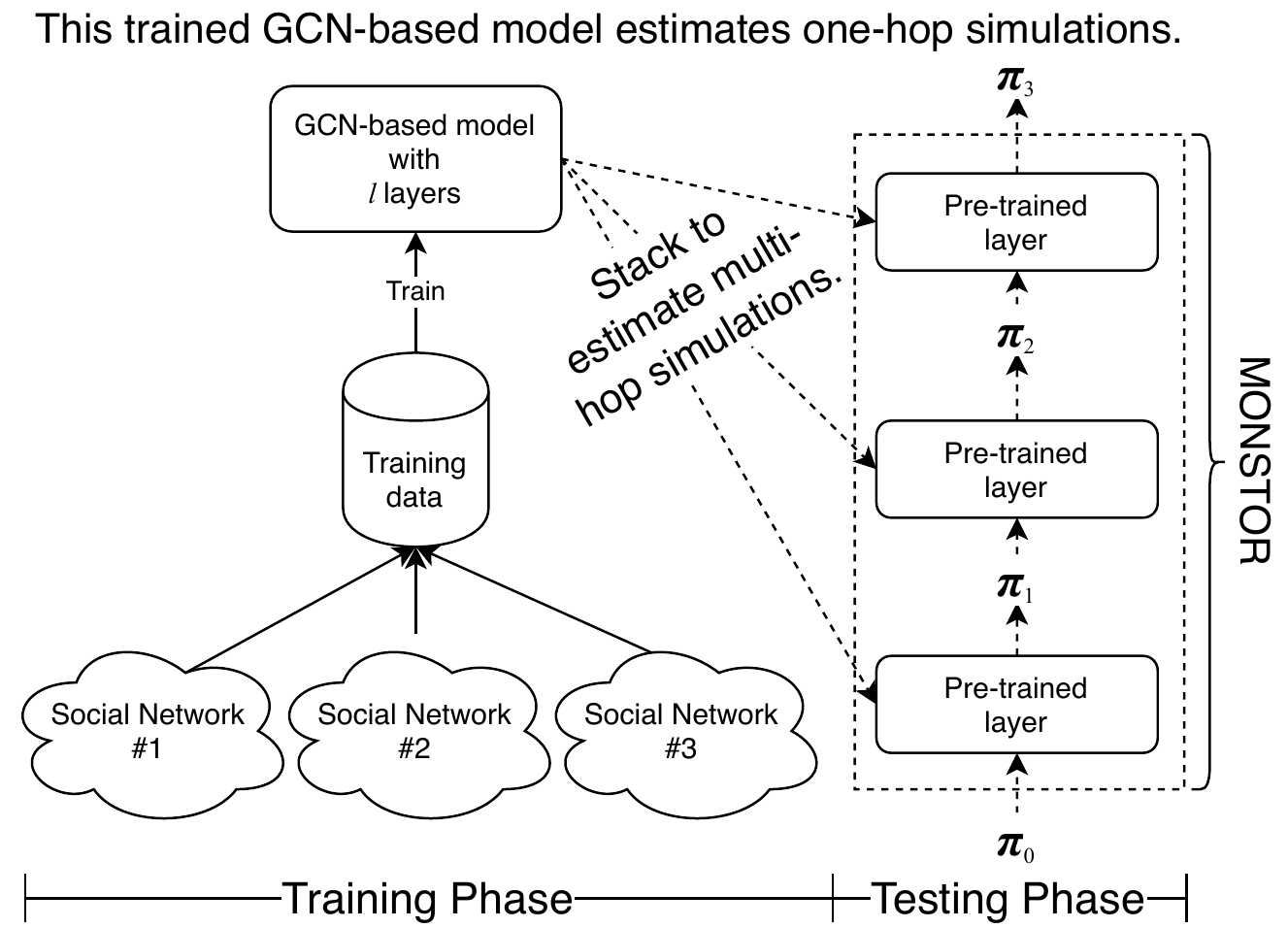}
    \caption{The overall workflow in our approach. We train a GCN and stack it $s$ (e.g., $3$ in this figure) times
    for testing. MONSTOR estimates $\boldsymbol{\pi}_3$ from $\boldsymbol{\pi}_0$.}
    \label{fig:overall}
\end{figure}

The overall workflow in
MONSTOR is as follows:
\begin{enumerate}[leftmargin=*]
    \item We collect one or more social networks $\{\GG_1, \GG_2, \cdots\}$.
    \item From each $\GG_j$, we collect the tuples
    $\{(\boldsymbol{\pi}_i,\boldsymbol{\pi}_{i-1},\cdots,\boldsymbol{\pi}_{i-e}, \allowbreak\mathbf{P}_j):i \geq e\}$, where $e>1$ is a hyperparameter, after choosing a seed set $S$ randomly so that $1 \leq |S| \leq \frac{|\V|}{50}$. $\mathbf{P}_j$ can be in BT, JI, or LP. We repeat this multiple times with different seed sets, as shown in Fig.~\ref{fig:sims}.
    \item We train our graph convolutional network (GCN)-based model $M$ with the training data.
    It has $l$ graph convolutional layers, and it estimates $\boldsymbol{\pi}_i$ given $\boldsymbol{\pi}_{i-1},\cdots,\boldsymbol{\pi}_{i-e}$,
    That is, $M$ estimates a single step of the IC model.
    \item We stack $s$ times the pre-trained model $M$, and this stacked model estimates $\boldsymbol{\pi}_s$ from $\boldsymbol{\pi}_0$. That is, it estimates end-to-end simulations under the IC model. Hereinafter, MONSTOR means the stacked GCN, described in Fig.~\ref{fig:overall}, unless otherwise stated. 
    \item For the IE problem, we compute $\langle 1,\boldsymbol{\pi}_s \rangle$ using the estimated $\boldsymbol{\pi}_s$. For the IM problem, we replace the MC simulation subroutine of existing IM algorithms (e.g., \cite{Kempe:2003:MSI:956750.956769,Goyal:2011:COG:1963192.1963217,6729575}) with MONSTOR.
\end{enumerate}
Note that, in the training phase, we use MC simulations to obtain 
$\boldsymbol{\pi}_i,\boldsymbol{\pi}_{i-1},\cdots,\boldsymbol{\pi}_{i-e}$. In the test phase, MC simulations in (potentially unseen) target networks are not needed. 

\subsection{Detailed Design}
We describe our GCN-based model $M$ and the training method for it. 
As stated earlier, $M$ estimates $\boldsymbol{\pi}_i$ given $\boldsymbol{\pi}_{i-1}$, $\cdots$, $\boldsymbol{\pi}_{i-e}$.
Specifically, $M$ initializes the feature vector of each node $v$ as $\mathbf{h}^0_{v}:=(\rho_{i-e+1}(v)-\rho_{i-e}(v),\cdots,\rho_{i-1}(v)-\rho_{i-2}(v), \rho_{i-1}(v))$, and it repeatedly computes new feature vectors of each node as follows  for $1 \leq i \leq l$:
\begin{small}
\begin{align*}
    &\mathbf{a}^i_v := \textsf{MAX}(\{p_{(u,v)} \cdot (\mathbf{h}^{i-1}_u\mathbf{W}^{i}_{1} + \mathbf{b}^{i}_{1})  :  u \in \textsf{NEI}(v)\}), \ \forall v \in V, \label{eq:ouragr} \\
     &\mathbf{h}^i_v := \textsf{ReLU}(\textsf{CONCAT}(\mathbf{h}^{i-1}_v, \mathbf{a}^{i}_v)\mathbf{W}^{i}_{2} + \mathbf{b}^{i}_{2}), \ \forall v \in V,
\end{align*}
\end{small}
\noindent where  $\mathbf{h}^{i}_{u}\in \mathbb{R}^{d_{i}}$ is the feature vector of the node $u$ at the $i$-th layer; $\textsf{NEI}(v)$ is the set of neighbors of $v$; \textsf{MAX} is the element-wise max function; \textsf{CONCAT} is the concatenation function;
and $\mathbf{W}^{i}_{1}\in \mathbb{R}^{d_{i-1} \times d_{i-1}}$, $\mathbf{W}^{i}_{2}\in \mathbb{R}^{ 2d_{i-1} \times d_{i}}$, $\mathbf{b}^{i}_{1}\in \mathbb{R}^{d_{i-1}}$ and $\mathbf{b}^{i}_{2}\in \mathbb{R}^{d_{i}}$ are learnable parameters.

The idea of multiplying $p_{(u,v)}$ and $\mathbf{h}^{i-1}_u$ is inspired by the fact that activation probabilities in the IC model are multiplied following a cascade route. For instance, the probability that $u_1$ activates $u_2$ and $u_2$ activates $u_3$ is $p_{(u_1,u_2)}\cdot p_{(u_2,u_3)}$.

Instead of directly estimating the raw values in $\boldsymbol{\pi}_i$, our model $M$ uses the following more effective estimation method, which is inspired by the monotonicity (see Proposition~\ref{pro:mono}):
\begin{small}\begin{equation}
    M(\boldsymbol{\pi}_{i-1},\cdots,\boldsymbol{\pi}_{i-e}, \mathbf{P};\bm{\theta}) := \min\{\boldsymbol{\pi}_{i-1} + \mathbf{h}^{l}, \mathbf{u}_{i}\},\label{eq:pos}
\end{equation}\end{small}
\noindent where $\bm{\theta}$ is the learnable parameters of $M$; $\mathbf{u}_{i}\in \mathbb{R}^{|\V|}$ is a theoretical upper bound of $\boldsymbol{\pi}_{i}$ (see Eq.~\eqref{eq:upper} below); and $\mathbf{h}^{l} \in \mathbb{R}^{|\V|}$ is the vector concatenating $\mathbf{h}^{l}_{v}\in \mathbb{R}$ (i.e., $d_{l}=1$) for all $v\in V$.
By Eq.~\eqref{eq:pos}, $\boldsymbol{\pi}_{i}$ always lies between $\boldsymbol{\pi}_{i-1}$ and $\mathbf{u}_{i}$, which are lower and upper bounds of $\boldsymbol{\pi}_{i}$. By adding these bounds, we relieve the difficulty of the estimation task. In our preliminary studies, directly estimating the raw values in $\boldsymbol{\pi}_i$ was not as successful as the proposed way.  



Of many possible loss functions, 
we train our model $M$ using the following loss function:
\begin{small}\begin{align*} 
    \mathcal{L} := \frac{1}{|T|} \sum_{t \in T} \left( \frac{ \| M(t;\bm{\theta}) - \boldsymbol{\pi}_i \|_1}{|\V|} + \lambda  \frac{|\mathbf{1}\cdot M(t;\bm{\theta}) - \langle \mathbf{1}, \boldsymbol{\pi}_i \rangle|}{\langle \mathbf{1}, \boldsymbol{\pi}_i \rangle}\right),
\end{align*}\end{small}
\noindent where $T$ is a training set; $t = (\boldsymbol{\pi}_{i},\boldsymbol{\pi}_{i-1},\cdots,\boldsymbol{\pi}_{i-e}, \mathbf{P}) \in T$ is a training sample; and $\lambda$ is a hyperparameter.
We aim to fit both individual infection probabilities and overall influence. 



\smallsection{Upper Bound of Infection Probabilities}
We prove the upper bound $\mathbf{u}_{i}$ of $\boldsymbol{\pi}_{i}$, which our estimation relies on. 

\vspace{1mm}
\noindent\fbox{%
    \parbox{\columnwidth}{%
        \begin{theorem}\label{thm:ub}
For all $i\geq 2$, the vector $\mathbf{u}_{i}$, defined as, 
\begin{small}
\begin{equation}
    \mathbf{u}_{i} := \boldsymbol{\pi}_{i-1} + (\boldsymbol{\pi}_{i-1} - \boldsymbol{\pi}_{i-2})\mathbf{P}, \label{eq:upper}
\end{equation}
\end{small}
is an upper bound of $\boldsymbol{\pi}_i$. That is, $\boldsymbol{\pi}_{i} \leq \mathbf{u}_{i}$. 
\end{theorem}
    }%
}

\vspace{1mm}

\textit{Proof:}
For each node $v\in \V$, we let $t_v$ be the step at which $v$ gets infected under the IC model. If $v$ is never infected, $t_v=\infty$.
For each node $v\in \V$ and its neighbor $u$, let $X_{u\rightarrow v}^{i}$ be the event that $v$ is infected by $u$ at step $i$. Then,
\begin{small}\begin{align}
    \mathbb{P}(t_{v} =  i) & = \mathbb{P}(\bigcup\nolimits_{u\in \textsf{NEI}(v)}X_{u\rightarrow v}^{i}) \ \leq \sum\nolimits_{u\in \textsf{NEI}(v)}\mathbb{P}(X_{u\rightarrow v}^{i}). \label{eq:prob:bound}
\end{align}\end{small}
In Eq.~\eqref{eq:prob:bound}, $\mathbb{P}(X_{u\rightarrow v}^{i}) = \mathbb{P}(t_{u} = i-1)\cdot p_{(u,v)}$ and $\mathbb{P}(t_{u} = i-1)=\rho_{i-1}(u)-\rho_{i-2}(u)$. Therefore, the following inequality holds for all $v\in \V$:
\begin{small}\begin{align*}
  \mathbb{P}&(t_{v}\leq i) = \mathbb{P}(t_{v}\leq i-1) + \mathbb{P}(t_{v} =  i) \\ 
  & \leq \mathbb{P}(t_{v}\leq i-1) + \sum\nolimits_{u\in \textsf{NEI}(v)} (\rho_{i-1}(u)-\rho_{i-2}(u))\cdot p_{(u,v)}
\end{align*}\end{small}
From Definition~\ref{defn:limited}, $\mathbb{P}(t_{v}\leq i)=\rho_{i}(v)$ and $\mathbb{P}(t_{v}\leq i-1)=\rho_{i-1}(v)$ hold, and these imply $\boldsymbol{\pi}_{i} \leq \mathbf{u}_{i}$. \hfill $\blacksquare$

\smallsection{Complexity and Runtime Analysis}
Once MONSTOR is trained \kijung{(potentially using graphs smaller than a target graph $\GG$)}, estimating $\boldsymbol{\pi}_{i}$ in $\GG=(\V,\E)$ for a seed set $S$ takes $\mathcal{O}(ls|\E|)$ time, where $s$ is the number of stacks, and $l$ is the number of convolutional layers per stack. In our experiments, the runtime of an estimation by MONSTOR amounts to the runtime of performing MC simulations only $100$ times. According to the standard configurations \cite{Kempe:2003:MSI:956750.956769,6729575}, Greedy and UBLF perform  MC simulations $10,000$ times per seed set.


\begin{table*}[t]
\centering
\caption{Statistics of each social network}\label{tbl:stat}
    \begin{tabular}{|c|c|c|c|c|c|c|c|c|}
    \hline
     & $|\V|$ & $|\E|$ & \multicolumn{2}{c|}{$\frac{\sum p_{(u,v)}}{|\E|}$ in BT} & \multicolumn{2}{c|}{$\frac{\sum p_{(u,v)}}{|\E|}$ in JI} & \multicolumn{2}{c|}{$\frac{\sum p_{(u,v)}}{|\E|}$ in LP} \\ \cline{4-9}
     & & & Train & Test & Train & Test & Train & Test \\ \hline
    Extended  & 11,409    & 58,972  & 0.07974  & 0.09194 & 0.03345 & 0.04095 & 0.16138 & 0.18371 \\ \hline
    WannaCry  & 35,627 & 169,419 & 0.07255 &    0.09466 & 0.02977 & 0.04494 & 0.19785 & 0.16297 \\ \hline
    Celebrity & 15,184 & 56,538 & 0.03206 & 0.02787 & 0.00163 & 0.00159 & 0.26142 & 0.256   \\ \hline
    \end{tabular}
\end{table*}






\begin{table*}[t]
\centering
\caption{\label{tbl:result2}\kijung{The accuracy (i.e., influence of output seeds) of IM methods. We highlight the accuracy in unseen datasets in \color{blue}{blue}.}
}
\subfloat[Test with BT]{
\begin{tabular}{|c|C{0.78cm}|C{0.78cm}|C{0.78cm}|C{0.78cm}|C{0.78cm}|C{0.78cm}|C{0.78cm}|C{0.78cm}|C{0.78cm}|}
\hline
& \multicolumn{3}{c|}{Extended} & \multicolumn{3}{c|}{WannaCry} & \multicolumn{3}{C{0.78cm}|}{Celebrity} \\ \cline{2-10}
& $k$=$10$ & $50$ & $100$ & $10$ & $50$ & $100$ & $10$ & $50$ & $100$ \\ \hline
Target Influence & 481.5 & 968.3 & 1222.9 & 991.6 & 2123.5 & 2752.3 & 52.8 & 105.3 & 155.2 \\ \hline
U-MON (E+W) & \textbf{481.1} & 966.8 & \textbf{1222.3} & 991.0 & 2122.8 & 2745.4 & {\color{blue}\textbf{52.7}} & {\color{blue}104.1} & {\color{blue}154.3} \\ \hline
U-MON (E+C) & 480.7 & 967.1 & 1221.9 & {\color{blue}\textbf{991.1}} & {\color{blue}2123.1} & {\color{blue}2744.8} & 52.6 & 104.2 & 154.5 \\ \hline
U-MON (W+C) & {\color{blue}\textbf{481.1}} & {\color{blue}\textbf{967.7}} & {\color{blue}1221.7} & 991.0 & \textbf{2123.2} & 2745.6 & 51.8 & \textbf{104.8} & \textbf{155.1} \\ \hline
D-SSA & 467.5 & 949.0 & 1189.8 & 984.2 & 2071.9 & 2687.1 & 49.8 & 102.0 & 152.4 \\ \hline 
SSA & 467.5 & 949.1 & 1189.8 & 984.3 & 2071.9 & 2687.2 & 49.8 & 102.0 & 152.4 \\ \hline
IRIE & 479.6 & 966.0 & 1221.7 & 986.6 & 2118.4 & \textbf{2751.8} & 51.7 & 103.0 & 153.0 \\ \hline
PMIA & 473.5 & 960.0 & 1199.5 & 989.0 & 2106.8 & 2739.9 & 51.7 & 100.0 & 152.1 \\ \hline
\end{tabular}

}
\\
\subfloat[Test with JI]{
\begin{tabular}{|c|C{0.78cm}|C{0.78cm}|C{0.78cm}|C{0.78cm}|C{0.78cm}|C{0.78cm}|C{0.78cm}|C{0.78cm}|C{0.78cm}|}
\hline
 & \multicolumn{3}{c|}{Extended} & \multicolumn{3}{c|}{WannaCry} & \multicolumn{3}{c|}{Celebrity} \\ \cline{2-10}
& $k$=$10$ & $50$ & $100$ & $10$ & $50$ & $100$ & $10$ & $50$ & $100$ \\ \hline
Target Influence  & 244.4 & 529.3 & 706.6 & 533.9 & 1238.5 & 1648.0 & 43.7 & 90.3 & 140.2 \\ \hline
U-MON (E+W) & 244.5 & 529.1 & 706.8 & \textbf{534.2} & \textbf{1239.2} & 1647.4 & {\color{blue}43.7} & {\color{blue}90.4} & {\color{blue}140.4} \\ \hline
U-MON (E+C) & 244.5 & \textbf{529.2} & 706.6 & {\color{blue}\textbf{534.2}} & {\color{blue}1239.1} & {\color{blue}1647.4} & 43.7 & \textbf{90.5} & 140.4 \\ \hline
U-MON (W+C) & {\color{blue}244.4} & {\color{blue}529.1} & {\color{blue}706.8} & 534.1 & 1239.1 & 1647.3 & 43.7 & 90.4 & \textbf{140.5} \\ \hline
D-SSA & 244.5 & 521.2 & 682.8 & 532.3 & 1213.6 & 1589.7 & 43.6 & 89.9 & 139.9 \\ \hline 
SSA & 244.5 & 521.2 & 682.8 & 532.3 & 1213.7 & 1589.8 & 43.6 & 89.9 & 139.9 \\ \hline
IRIE & 244.3 & \textbf{529.2} & \textbf{707.3} & \textbf{534.2} & 1239.0 & \textbf{1647.8} & \textbf{43.8} & 90.4 & 140.3 \\ \hline
PMIA & \textbf{244.6} & 529.1 & 705.4 & 534.1 & 1239.0 & 1646.8 & 42.7 & 90.4 & 140.3 \\ \hline
\end{tabular}
}
\\
\subfloat[Test with LP]{
\begin{tabular}{|c|C{0.78cm}|C{0.78cm}|C{0.78cm}|C{0.78cm}|C{0.78cm}|C{0.78cm}|C{0.78cm}|C{0.78cm}|C{0.78cm}|}
\hline
& \multicolumn{3}{c|}{Extended} & \multicolumn{3}{c|}{WannaCry} & \multicolumn{3}{c|}{Celebrity} \\ \cline{2-10}
& $k$=$10$ & $50$ & $100$ & $10$ & $50$ & $100$ & $10$ & $50$ & $100$ \\ \hline
Target Influence  & 1852.4 & 2876.5 & 3264.9 & 5271.6 & 7880.0 & 9098.3 & 5508.4 & 5616.7 & 5657.7 \\ \hline
C-MON (E+W) & 1843.0 & \textbf{2863.0} & \textbf{3253.8} & 5246.4 & \textbf{7862.1} & \textbf{9073.5} & {\color{blue}5508.9} & {\color{blue}5615.0} & {\color{blue}5664.9} \\ \hline
C-MON (E+C) & 1840.6 & 2848.5 & 3236.5 & {\color{blue}5253.0} & {\color{blue}7844.5} & {\color{blue}9041.7} & 5508.8 & 5616.4 & 5666.4 \\ \hline
C-MON (W+C) & {\color{blue}1839.5} & {\color{blue}2853.1} & {\color{blue}3242.0} & 5248.6 & 7850.7 & 9045.7 & 5508.8 & 5615.0 & 5665.0 \\ \hline
D-SSA & \textbf{1844.3} & 2858.7 & 3236.1 & 5256.7 & 7783.4 & 8977.3 & 5509.0 & 5606.2 & 5633.8 \\ \hline 
SSA & 1843.8 & 2858.6 & 3236.1 & \textbf{5257.2} & 7783.6 & 8977.0 & 5508.8 & 5606.3 & 5633.9 \\ \hline
IRIE & 1816.2 & 2829.8 & 3201.2 & 5109.1 & 7714.1 & 8840.1 & \textbf{5509.1} & \textbf{5617.4} & \textbf{5667.4} \\ \hline
PMIA & 1830.0 & 2828.9 & 3243.2 & 5196.7 & 7807.6 & 8981.8 & 5508.5 & 5604.2 & 5630.2 \\ \hline
\end{tabular}
}
\end{table*}

\section{Experiments}
\label{sec:exp}

We review our experiments for answering the following questions:
\begin{itemize}
    \item \textbf{Q1. Influence Maximization}: How accurate are simulation-based IM algorithms equipped with MONSTOR, compared to state-of-the-art competitors?
    \item \textbf{Q2. Influence Estimation}: How accurately does MONSTOR estimate the influence of seed sets? 
    \item \textbf{Q3. Scalability}: How rapidly does the estimation time grow as the size of the input graph increase?
    \item \textbf{Q4. Submodularity}: Is MONSTOR submoudular as the ground-truth influence function is?
\end{itemize}
We made the datasets, the source code, and the trained models available at \textsf{\url{https://github.com/jihoonko/asonam20-monstor}}.

\subsection{Experimental Settings}
\label{sec:exp:setting}
\smallsection{Datasets} We used three real-world social networks: Extended, WannaCry, and Celebrity (see Table~\ref{tbl:stat}). For Extended, we crawled more tweets and retweets in addition to those used in~\cite{191006}.
In each dataset, we used online postings (and their cascade logs) 
during the first 50\% of time for training/validation and those during the remaining 50\% for testing --- two sets are disjoint. Specifically, we computed the activation probability matrices $\mathbf{P}_{BT}$, $\mathbf{P}_{JI}$, and $\mathbf{P}_{LP}$ for each of the training/validation and testing periods based on the logs. Then, we collected $1,600$ training tuples, $400$ validation tuples and $2,000$ testing tuples, as described in Sec.~\ref{sec:over}.

\begin{figure}[t]
    \centering
    \includegraphics[width=\columnwidth]{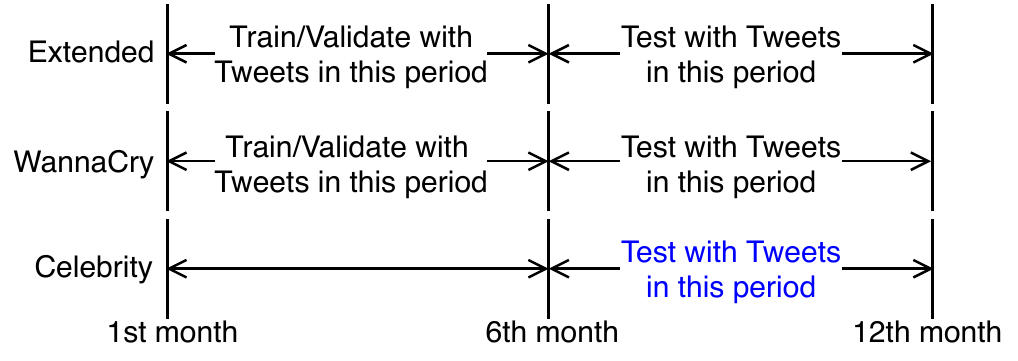} \\
    \caption{Example train-test split in our experiments. We test with tweets newly posted in the test period.
    Note that MONSTOR can be applied to unseen social networks (i.e., the test data highlighted in blue), while existing learning approaches~\cite{DBLP:conf/atal/YanSLMLS19,DBLP:journals/corr/abs-1906-07378} cannot be applied.}
    \label{fig:split}
\end{figure}

We trained MONSTOR with training data consisting of two out of the three networks (e.g., Extended and WannaCry) and tested it with each of the three networks (i.e., each of Extended, WannaCry, and Celebrity), as shown in Fig.~\ref{fig:split}.



\smallsection{Competitors} We consider the following IM algorithms as competitors: (i) Greedy~\cite{Kempe:2003:MSI:956750.956769}, UBLF~\cite{6729575}, and CELF~\cite{Goyal:2011:COG:1963192.1963217} among simulation-based algorithms; (ii) SSA~\cite{10.1145/2882903.2915207}, D-SSA~\cite{10.1145/2882903.2915207}, PMIA~\cite{pmia}, and IRIE~\cite{DBLP:journals/corr/abs-1111-4795} among non-simulation-based algorithms; and  (iii) U-MON and C-MON, where UBLF and CELF equipped with MONSTOR which replaces MC simulations. We note that Greedy, UBLF, and CELF output the same seed users. \kijung{Among them, UBLF is the fastest, and CELF and Greedy follow.} U-MON and C-MON also output the same seed users. UBLF and U-MON can be used only when activation probabilities satisfy a certain property~\cite{6729575}.


\smallsection{Hyperparameters} 
For MONSTOR, we set $e$=$4$, $l$=$3$, $\lambda$=$0.3$, $d_{1}$=$\cdots$= $d_{l-1}$=$16$ after some preliminary studies.
At each $t$-th epoch, we set the learning rate to $10^{-4} \cdot t$ if $t\leq 10$, and  $10^{-2}/t$ otherwise.
\jihoon{We chose the best $s$ using the validation data, and $s=2,3,5$ were the best for JI, BT, and LP, respectively.} 
For (D-)SSA, we set $\epsilon$=$0.1$ and $\delta$=$1/|\V|$ as in~\cite{10.1145/2882903.2915207} and used the influence averaged over $100$ independent runs. For IRIE and PMIA, we followed the settings in~\cite{DBLP:journals/corr/abs-1111-4795,pmia}.



\begin{table}[t]
\centering
\caption{The accuracy of MONSTOR on the influence estimation problem. We highlight the accuracy in unseen datasets in \color{blue}{blue}.}
\label{tbl:ablation}
\subfloat[Extended]{
\scalebox{0.9}{
\hspace{-3.5mm}
\begin{tabular}{|c|c|c|c|c|c|c|}
    \hline
    & \multicolumn{3}{c|}{Pearson Correlation} & \multicolumn{3}{c|}{Spearman's Rank} \\ \cline{2-7}
    & BT & JI & LP & BT & JI & LP \\ \hline
    MON. (E+W) & 1.000 & 1.000 & 1.000 & 1.000 & 1.000 & 1.000 \\ \hline
    MON. (E+C) & 1.000 & 1.000 & 1.000 & 1.000 & 1.000 & 0.999 \\ \hline
    {\color{blue}MON. (W+C)} & 1.000 & 1.000 & 1.000 & 1.000 & 1.000 & 0.999 \\ \hline
    \end{tabular}
}}

\subfloat[WannaCry]{
\scalebox{0.9}{
\hspace{-3.5mm}
\begin{tabular}{|c|c|c|c|c|c|c|c|c|c|}
    \hline
    & \multicolumn{3}{c|}{Pearson Correlation} & \multicolumn{3}{c|}{Spearman's Rank} \\ \cline{2-7}
    & BT & JI & LP & BT & JI & LP \\ \hline
    MON. (E+W) & 1.000 & 1.000 & 1.000 & 1.000 & 1.000 & 1.000 \\ \hline
    {\color{blue}MON. (E+C)} & 1.000 & 1.000 & 1.000 & 1.000 & 1.000 & 1.000 \\ \hline
    MON. (W+C) & 1.000 & 1.000 & 1.000 & 1.000 & 1.000 & 1.000 \\ \hline
    \end{tabular}
}}

\subfloat[Celebrity]{
\scalebox{0.9}{
\hspace{-3.5mm}
\begin{tabular}{|c|c|c|c|c|c|c|c|c|c|}
    \hline
    & \multicolumn{3}{c|}{Pearson Correlation} & \multicolumn{3}{c|}{Spearman's Rank} \\ \cline{2-7}
    & BT & JI & LP & BT & JI & LP \\ \hline
    {\color{blue}MON. (E+W)} & 0.999 & 1.000 & 1.000 & 0.999 & 1.000 & 0.998 \\ \hline
    MON. (E+C) & 1.000 & 1.000 & 1.000 & 1.000 & 1.000 & 0.999 \\ \hline
    MON. (W+C) & 1.000 & 1.000 & 1.000 & 1.000 & 1.000 & 0.998 \\ \hline
    \end{tabular}
}}
\end{table}

\begin{table}
    \centering
    \caption{Scatter plots for comparison of ground-truth influences and those estimated by MONSTOR.\label{tbl:ie_plots}}
    \subfloat[Test with BT]{
    \centering
    \begin{tabular}{c|ccc}
    & Extended & WannaCry & Celebrity \\ \hline
    \rotatebox[origin=l]{90}{\footnotesize{U-MON (E+W)} \ } &
    \includegraphics[height=0.74in]{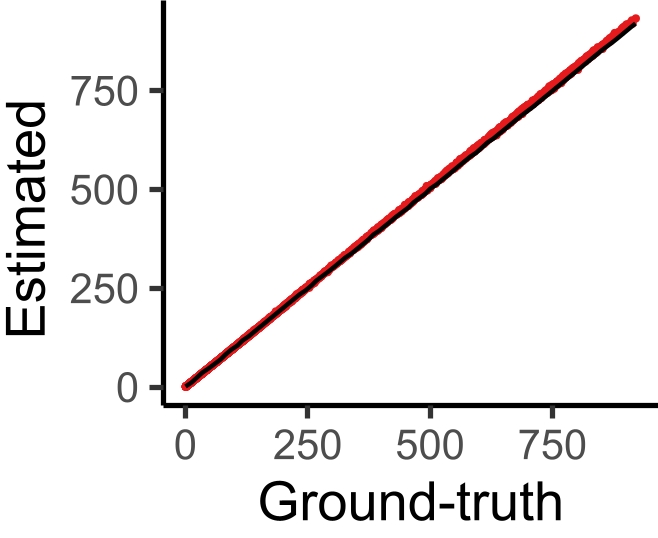} &
    \includegraphics[height=0.74in]{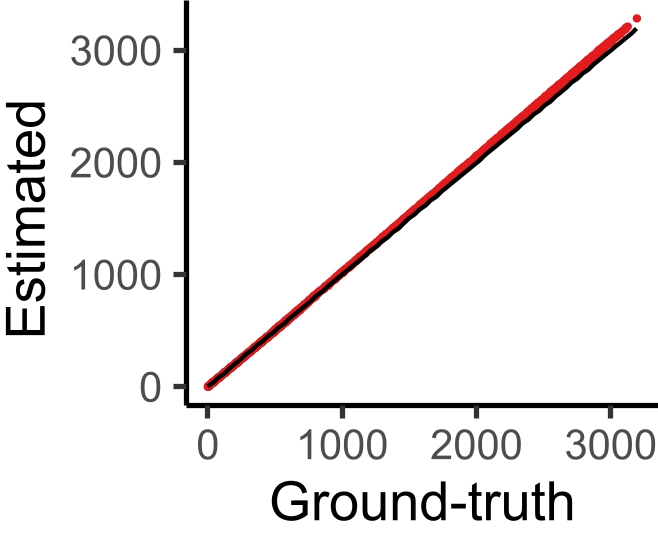} &
    \includegraphics[height=0.74in]{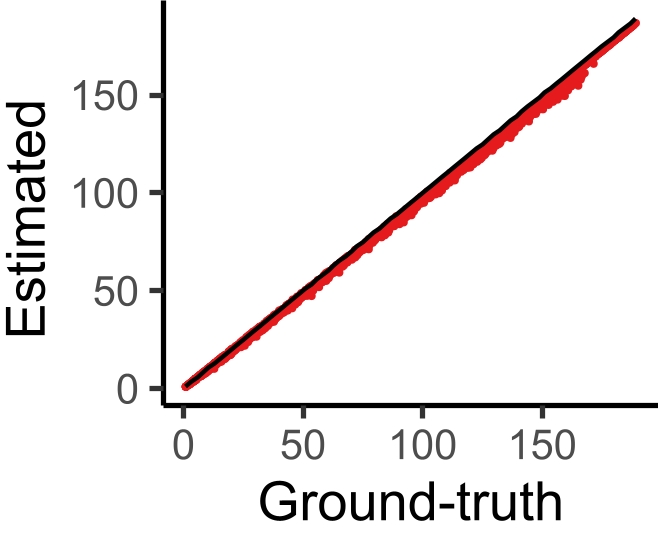} \\
    \rotatebox[origin=l]{90}{\footnotesize{U-MON (E+C)}} &
    \includegraphics[height=0.74in]{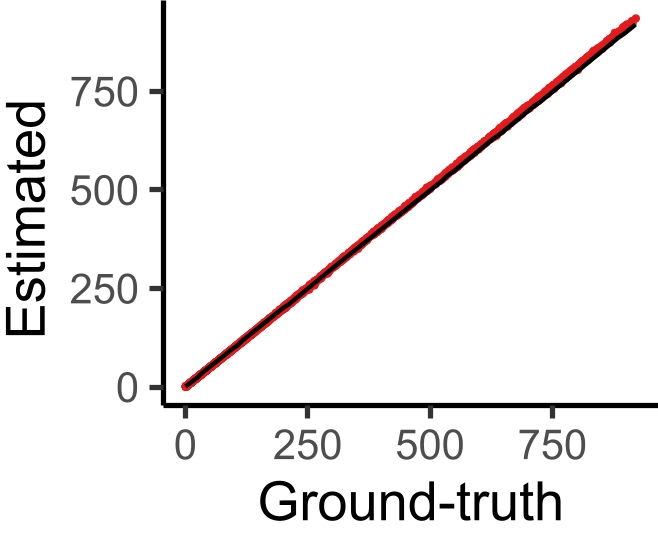} &
    \includegraphics[height=0.74in]{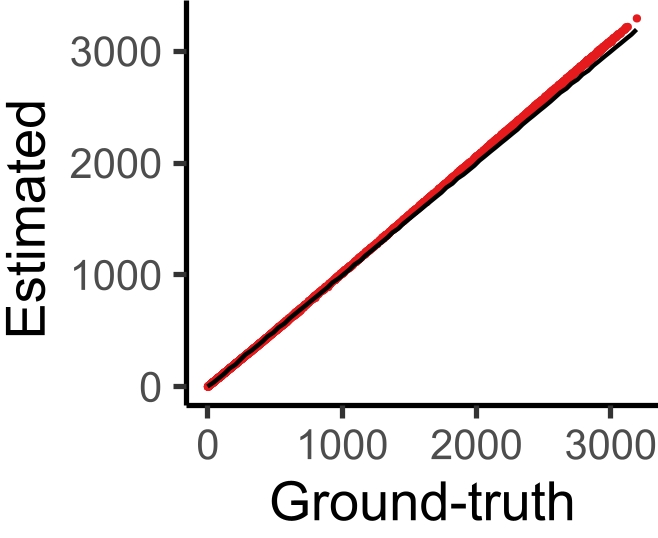} &
    \includegraphics[height=0.74in]{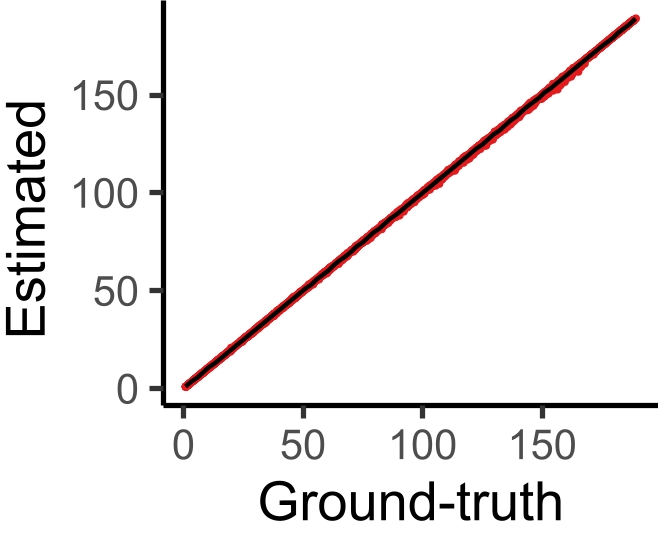} \\
    \rotatebox[origin=l]{90}{\footnotesize{U-MON (W+C)}} &
    \includegraphics[height=0.74in]{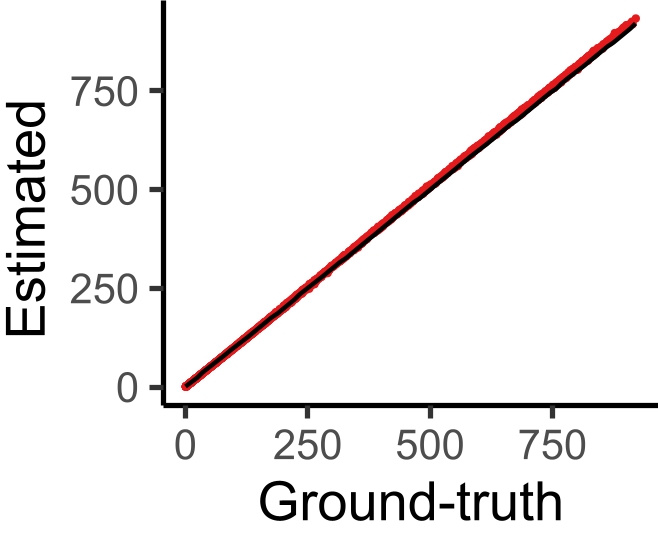} &
    \includegraphics[height=0.74in]{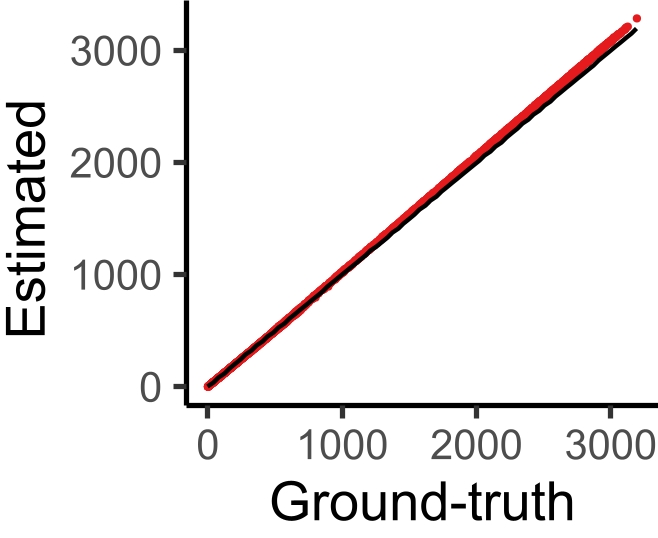} &
    \includegraphics[height=0.74in]{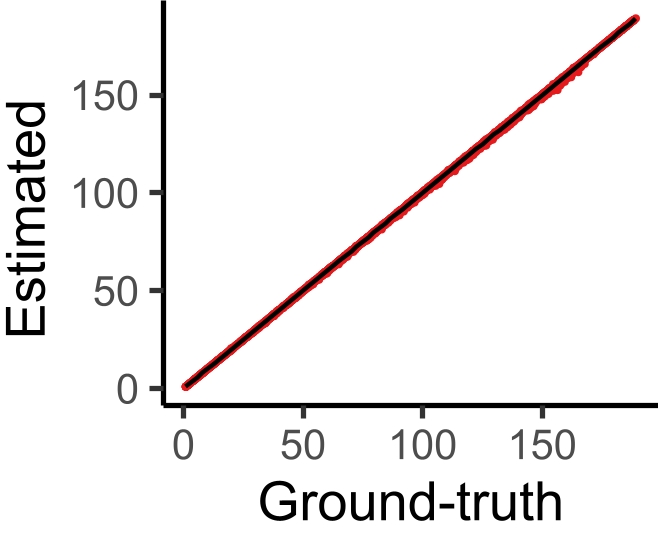} \\
    \end{tabular}}
    
    \subfloat[Test with JI]{
    \begin{tabular}{c|ccc}
    & Extended & WannaCry & Celebrity \\ \hline
    \rotatebox[origin=l]{90}{\footnotesize{U-MON (E+W)} \ } &
    \includegraphics[height=0.74in]{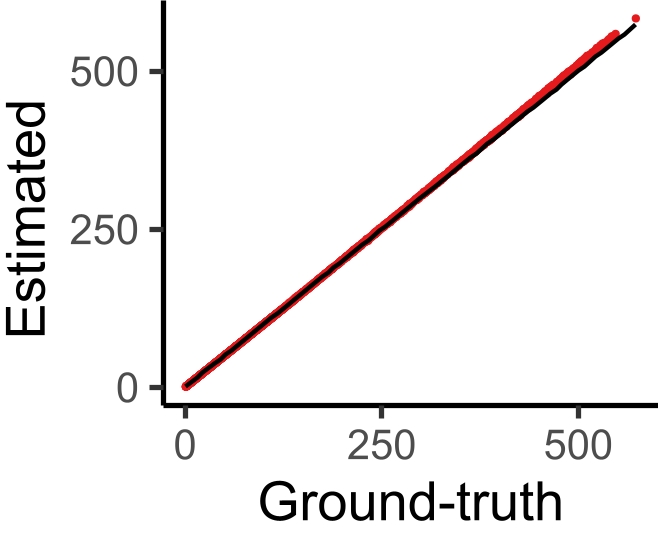} &
    \includegraphics[height=0.74in]{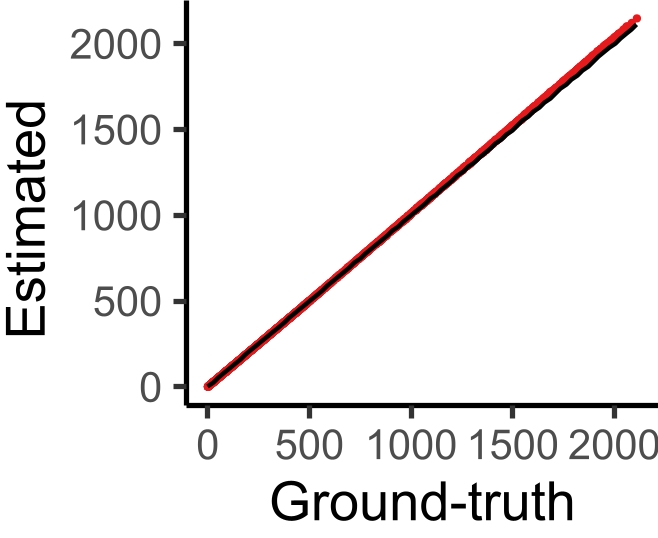} &
    \includegraphics[height=0.74in]{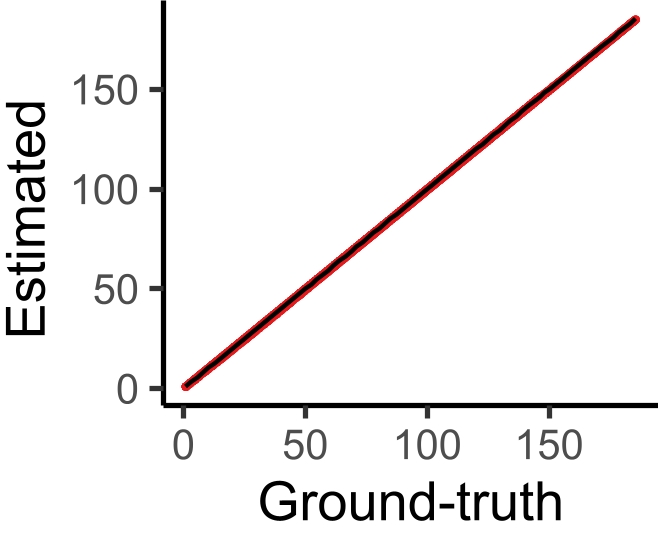} \\
    \rotatebox[origin=l]{90}{\footnotesize{U-MON (E+C)}} &
    \includegraphics[height=0.74in]{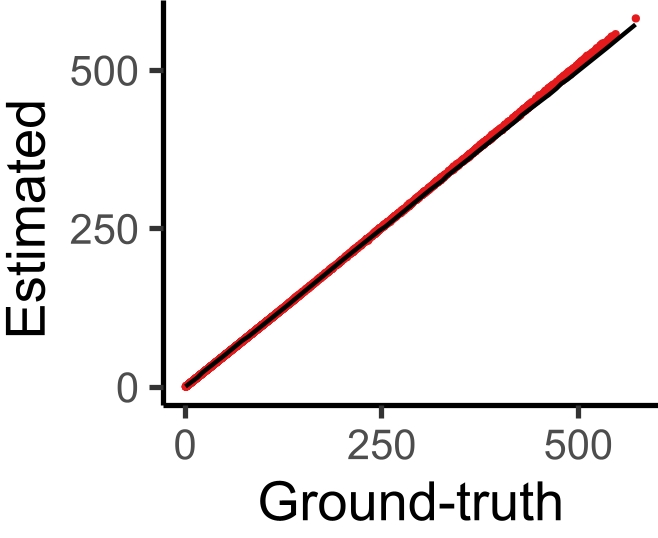} &
    \includegraphics[height=0.74in]{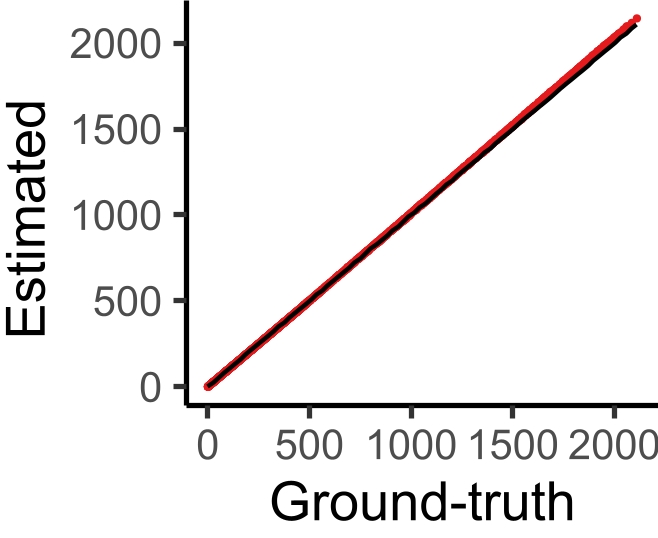} &
    \includegraphics[height=0.74in]{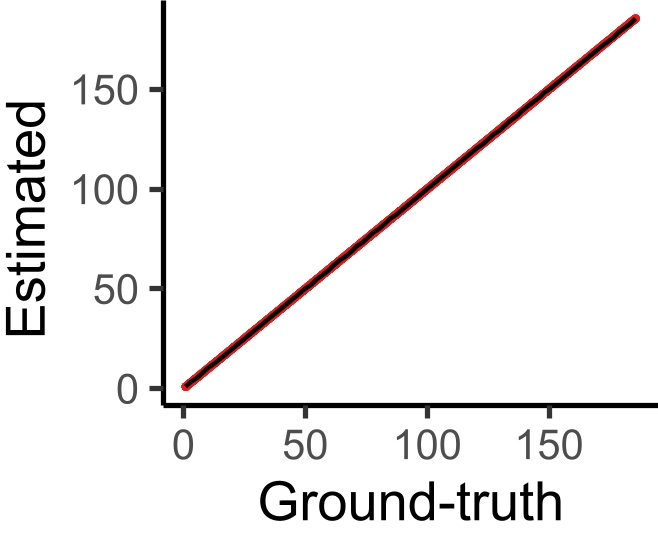} \\
    \rotatebox[origin=l]{90}{\footnotesize{U-MON (W+C)}} &
    \includegraphics[height=0.74in]{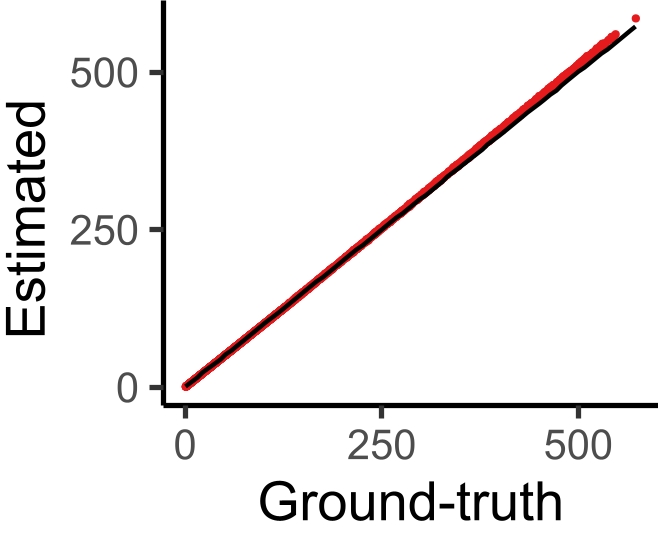} &
    \includegraphics[height=0.74in]{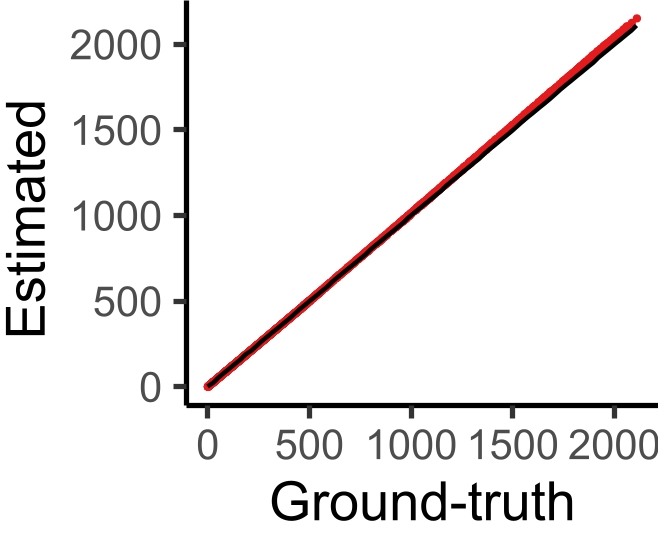} &
    \includegraphics[height=0.74in]{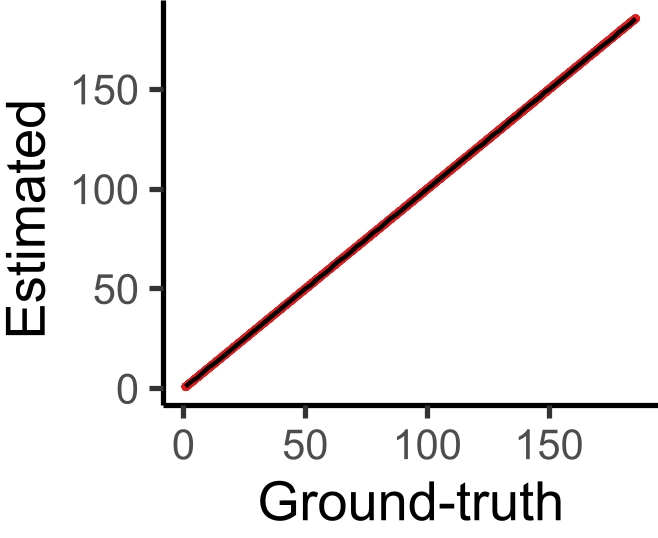} \\
    \end{tabular}}
    
    \subfloat[Test with LP]{
    \begin{tabular}{c|ccc}
    & Extended & WannaCry & Celebrity \\ \hline
    \rotatebox[origin=l]{90}{\footnotesize{C-MON (E+W)} \ } &
    \includegraphics[height=0.74in]{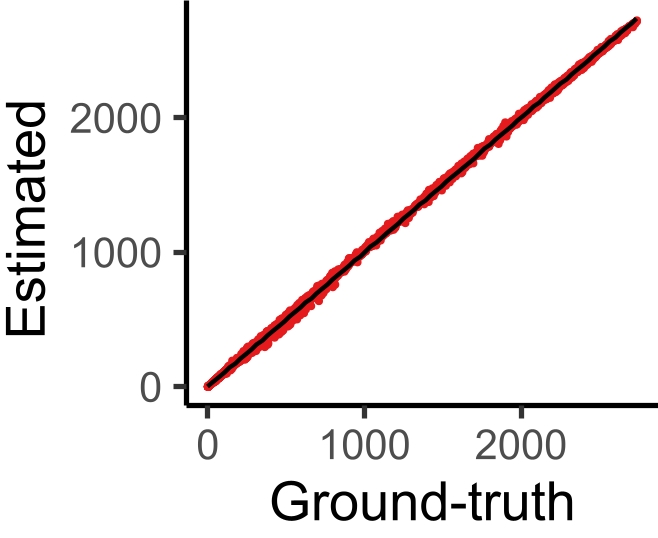} &
    \includegraphics[height=0.74in]{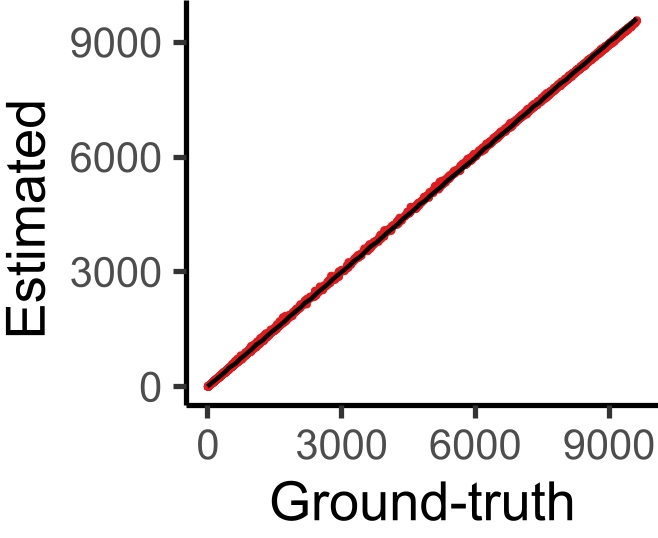} &
    \includegraphics[height=0.74in]{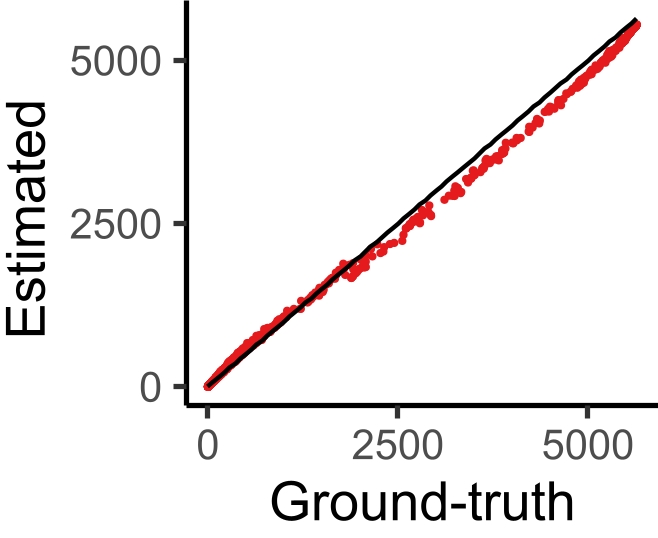} \\
    \rotatebox[origin=l]{90}{\footnotesize{C-MON (E+C)}} &
    \includegraphics[height=0.74in]{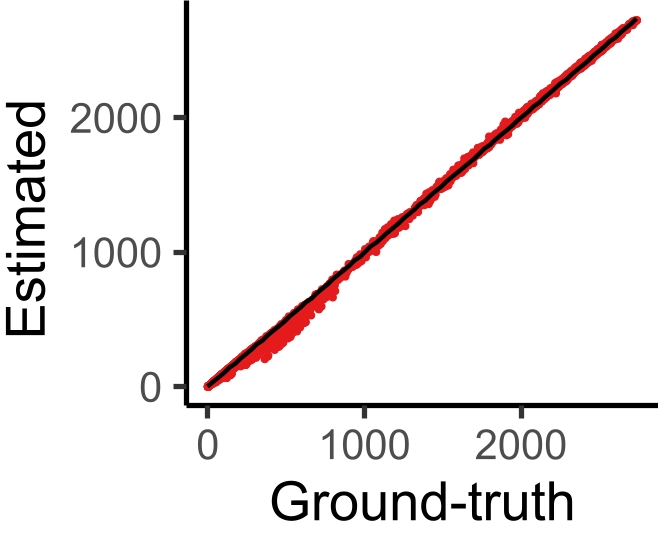} &
    \includegraphics[height=0.74in]{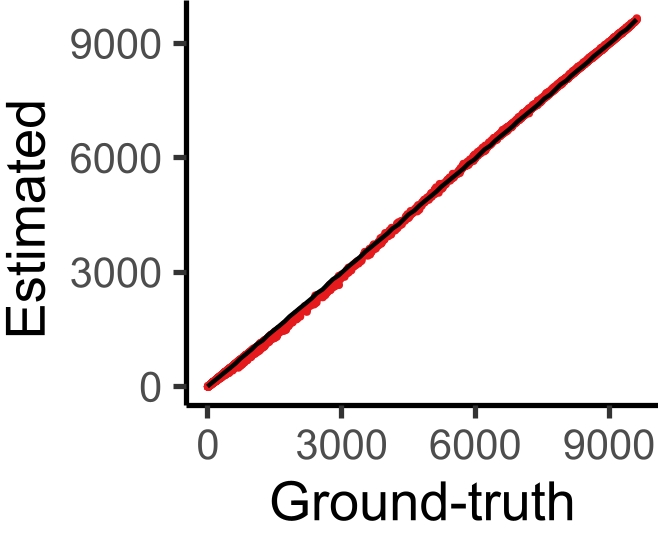} &
    \includegraphics[height=0.74in]{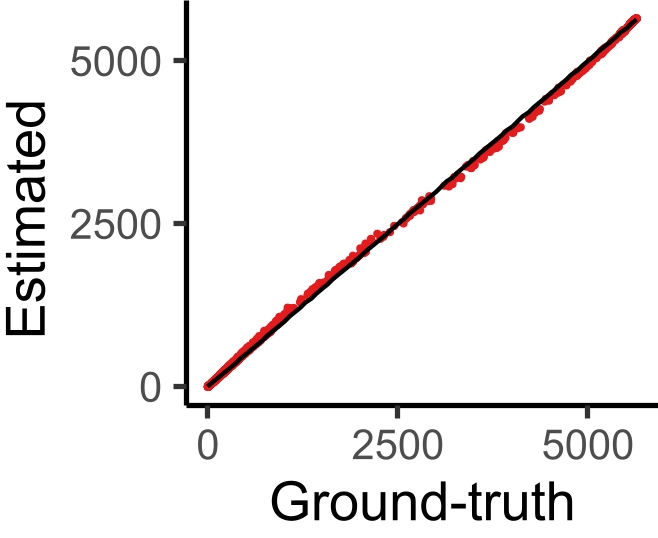} \\
    \rotatebox[origin=l]{90}{\footnotesize{C-MON (W+C)}} &
    \includegraphics[height=0.74in]{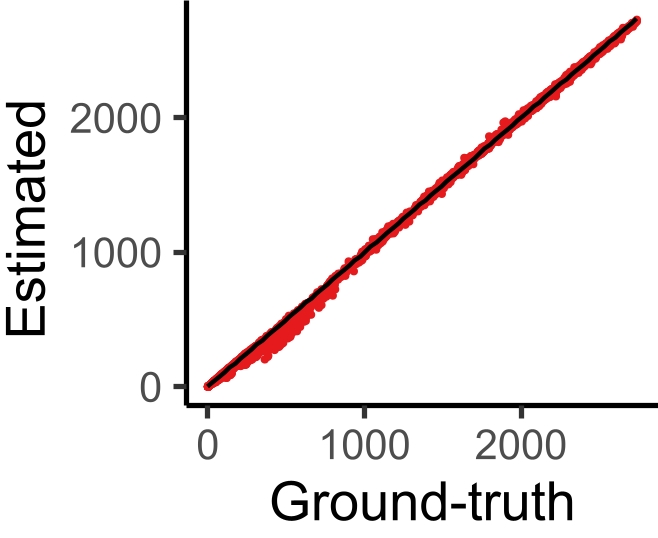} &
    \includegraphics[height=0.74in]{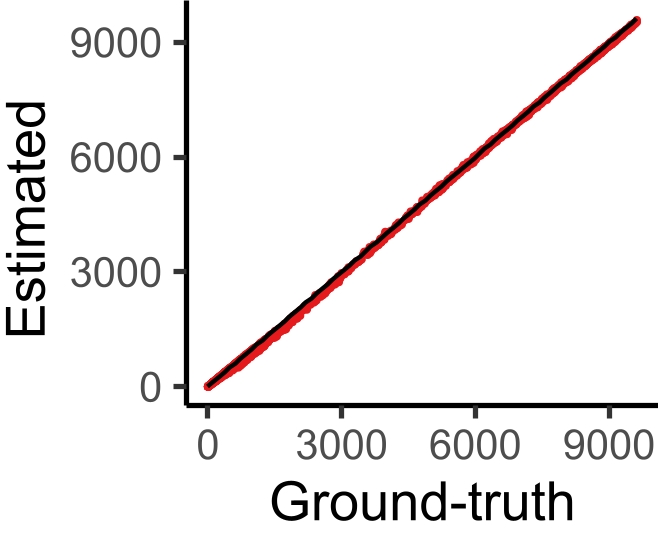} &
    \includegraphics[height=0.74in]{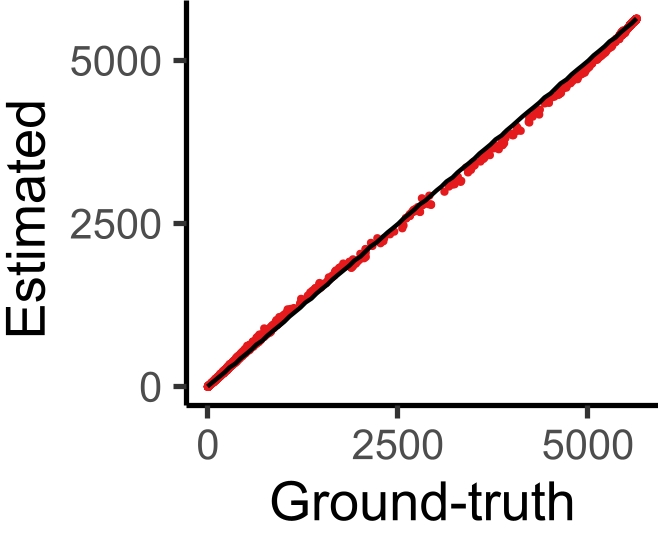} \\
    \end{tabular}}
\end{table}

\begin{table*}[t]
\centering
\caption{\label{tbl:time}The runtime of $1,000$ estimations by MONSTOR in graphs with different numbers of edges. 
}
    \begin{tabular}{|c|c|c|c|c|c|c|c|}
    \hline
    $|\E|$ & $2^{20}$ & $2^{21}$ & $2^{22}$ & $2^{23}$ & $2^{24}$ & $2^{25}$ &  $2^{26}$ \\ \hline
    Estimation time (sec) & 11.5 & 17.7 & 31.0 & 56.3 & 108.9 & 411.0 & 819.7 \\ \hline
    \end{tabular}
\end{table*}


\subsection{Q1. Influence Maximization (IM)}
We report the accuracy (i.e., influence of output seeds) of IM methods in Table~\ref{tbl:result2}.
Target influences mean influences obtained by Greedy, UBLF, and CELF with $10,000$ MC simulations, which took at least two orders of magnitude more time than the compared methods.
 

For BT, U-MON was most accurate in most cases. Its score was very close to the target influence.
IRIE was most accurate in one case. 
For JI, while U-MON and IRIE were most accurate in many cases, and there was no clear winner. Each algorithm (except SSA and D-SSA) was most accurate in at least one case. One possible reason for this is that activation probabilities are relatively small compared to BT and LP, as shown in Table~\ref{tbl:stat}.
 For LP, C-MON was most accurate in most cases, and its influence was very close to the target influence.
Note that the activation probabilities are relatively large in LP (see Table~\ref{tbl:stat}), and thus even small changes in the seed set can decrease influence significantly. 
Interestingly, IRIE was outperformed by D-SSA and SSA in many cases, while the results were the opposite when BT and JI were used.

In summary, only U-MON and C-MON were consistently accurate regardless of the activation probabilities. In most cases and even in unseen social networks, their influences were close to the target influence made by Greedy, UBLF and CELF with $10,000$ MC simulations. The other algorithms (i.e., D-SSA, SSA, IRIE, and PMIA) were not as accurate as our methods for at least one among BT, JI and LP.

\subsection{Q2. Influence Estimation (IE)} To explain how our methods yielded high quality seed sets, as shown in Table~\ref{tbl:result2}, we show that MONSTOR estimates the influence of candidate seed sets accurately.
For each test seed set described in Section~\ref{sec:exp:setting},
we estimated its influence using MONSTOR and used the mean influence of $10,000$ MC simulations as the ground-truth influence.
To measure the similarity between true and estimated influences over all test seed sets, we used Pearson's correlation coefficients and Spearman's Rank correlation coefficients. 
As seen in Table~\ref{tbl:ablation} and~\ref{tbl:ie_plots}, the ground-truth influences of test seed sets and the estimated influences are highly correlated, and both coefficients were close to $1.0$. That is, MONSTOR yielded near perfect estimates.

\subsection{Q3. Scalability} 
We demonstrate the linear scalability of MONSTOR.
To this end, we generated realistic graphs of various sizes using the R-MAT generator~\cite{chakrabarti2004r} with $a=0.7$ and $b=c=d=0.1$.
The number of edges in the generated graphs ranged from $2^{20}$ to $2^{26}$, and for every graph, we set the number of nodes to $20\%$ of the number of edges. 
We used the weighted cascade model~\cite{Kempe:2003:MSI:956750.956769} to determine the activation probability of each edge in the generated graphs.  
We measured the runtime of influence estimations by MONSTOR with $1,000$ different seed sets.
For each seed set, we chose its size uniformly at random from 1 to 10\% of $|\V|$ and then chose seed nodes uniformly at random. 
The runtime scaled linearly with the number of stacked GCNs (i.e., $s$), and we measured the runtime per stacked GCN.
In Table~\ref{tbl:time}, we report the runtime of $1,000$ influence estimations by MONSTOR in graphs of different sizes. 
Consistently with the theoretical result in Section~\ref{sec:method}, the runtime was near linear in the number of edges in the input graph.

\begin{table}[t]
\centering
\caption{The ratio of the cases where the submodularity holds.
We highlight the cases with unseen datasets in \color{blue}{blue}.
}\label{tbl:sub2}
\subfloat[Test with BT]{
\begin{tabular}{|c|c|c|c|}
\hline
 & Extended & WannaCry & Celebrity \\ \hline
U-MON (E+W) & 0.9996 & 0.9999 & {\color{blue}0.9999}  \\ \hline
U-MON (E+C) & 0.9994 & {\color{blue}0.9999} & 1.0000  \\ \hline
U-MON (W+C) & {\color{blue}0.9997} & 0.9999 & 1.0000  \\ \hline
\end{tabular}
}

\subfloat[Test with JI]{
\begin{tabular}{|c|c|c|c|}
\hline
 & Extended & WannaCry & Celebrity \\ \hline
U-MON (E+W) & 0.9988 & 0.9995 & {\color{blue}0.9966}  \\ \hline
U-MON (E+C) & 0.9993 & {\color{blue}0.9995} & 0.9999  \\ \hline
U-MON (W+C) & {\color{blue}0.9990} & 0.9994 & 0.9998  \\ \hline
\end{tabular}
}

\subfloat[Test with LP]{
\begin{tabular}{|c|c|c|c|}
\hline
 & Extended & WannaCry & Celebrity \\ \hline
C-MON (E+W) & 0.9993 & 0.9997 & {\color{blue}0.9938}  \\ \hline
C-MON (E+C) & 0.9994 & {\color{blue}0.9997} & 0.9970  \\ \hline
C-MON (W+C) & {\color{blue}0.9992} & 0.9996 & 0.9945  \\ \hline
\end{tabular}
}
\end{table}

\begin{table}[t]
\centering
\caption{MAPE when the submodularity does not hold.
We highlight the cases with unseen datasets in \color{blue}{blue}.
}\label{tbl:sub1}
\subfloat[Test with BT]{
\begin{tabular}{|c|c|c|c|}
\hline
 & Extended & WannaCry & Celebrity \\ \hline
U-MON (E+W) & 2.50e-04 & 5.40e-07 & {\color{blue}1.55e-04}  \\ \hline
U-MON (E+C) & 1.48e-05 & {\color{blue}4.42e-08} & 1.50e-04  \\ \hline
U-MON (W+C) & {\color{blue}7.67e-08} & 7.11e-08 & 8.02e-08  \\ \hline
\end{tabular}
}

\subfloat[Test with JI]{
\begin{tabular}{|c|c|c|c|}
\hline
 & Extended & WannaCry & Celebrity \\ \hline
U-MON (E+W) & 6.91e-06 & 5.60e-08 & {\color{blue}4.58e-05}  \\ \hline
U-MON (E+C) & 1.28e-06 & {\color{blue}4.69e-08} & 2.22e-05  \\ \hline
U-MON (W+C) & {\color{blue}3.93e-07} & 1.39e-07 & 1.10e-05  \\ \hline
\end{tabular}
}

\subfloat[Test with LP]{
\centering
\begin{tabular}{|c|c|c|c|}
\hline
 & Extended & WannaCry & Celebrity \\ \hline
C-MON (E+W) & 0.0022 & 0.0013 & {\color{blue}0.0114}  \\ \hline
C-MON (E+C) & 0.0025 & {\color{blue}0.0021} & 0.0113  \\ \hline
C-MON (W+C) & {\color{blue}0.0034} & 0.0011 & 0.0095  \\ \hline
\end{tabular}
}
\end{table}

\subsection{Q4. Submodularity}
It is well known that the influence maximization is a submodular maximization problem, and many IM algorithms exploit the submodularity of the ground-truth influence function. 
Therefore, it is crucial to show that MONSTOR has the same characteristic for claiming that IM algorithms based on the submodularity work properly when being integrated with MONSTOR.

In this section, we review our experiments for testing the empirical submodularity of MONSTOR.
To this end, we used $5,000$ test seed sets that were not used for training.
For each seed set, we chose its size uniformly at random from 1 to 10\% of $|\V|$ and then chose seed nodes uniformly at random. 
Using each pair $S$ and $T$ of the seed sets, we tested whether the following submodularity condition is met: 
$$f(S) + f(T) \geq f(S \cup T) + f(S \cap T).$$

In Table~\ref{tbl:sub2}, we show the ratio of the pairs where the above submodularity condition is met. When BT or JI was used, for more than 99.6\% of the pairs, the submodularity condition held. 
When LP was used, the ratio was $99\%$ or higher.

For each pair $S$ and $T$ where the submodularity condition was not met, we measured the MAPE (i.e., mean absolute percentage error) as follows:$$\frac{f(S \cup T) + f(S \cap T) - f(S) + f(T)}{f(S \cup T) + f(S \cap T)}.$$
As shown in Table~\ref{tbl:sub1}, the error (i.e., $f(S \cup T) + f(S \cap T) - f(S) + f(T)$) was marginal compared to the actual influence (i.e, $f(S \cup T) + f(S \cap T)$).
All these experiment results support that influence estimation by MONSTOR can be considered as submodular in practice.


\section{Related Work}
\label{sec:rel}
Influence maximization, i.e., to find a certain number of seed nodes who maximize the information spread through a social network, is an NP-hard problem. \textit{Independent cascade} (IC) and \textit{linear threshold} (LT) are the two most popular information cascade models.
In the IC model, once a node $u$ is activated (or influenced), it attempts once to activate each neighbor $v$ with probability $p_{(u,v)}$. 
In the LT model, a node $v$ is activated if a sufficient number of its neighbors (larger than a threshold) are activated.
In this work, we focus on the IC model. There are several successful real-world studies based on the IC model~\cite{Yadav:2016:USN:2937029.2937034,10.5555/3237383.3237428}.

Numerous methods have been proposed for influence maximization. Since it is an NP-hard problem, all these methods approximate optimal seed nodes. They can be categorized into the following three types: i) simulation-based, ii) proxy-based, and iii) sketch-based methods. Among them, simulation-based methods are known to be able to find better seeds than the other methods.

In the simulation-based methods, MC simulations are explicitly repeated to estimate the influence of seed sets~\cite{Goyal:2011:COG:1963192.1963217,6729575}. These methods focus on pruning unnecessary (redundant) simulations to minimize the required number of simulations. SSA~\cite{10.1145/2882903.2915207} and D-SSA~\cite{10.1145/2882903.2915207} are strong among sketch-based method, and IRIE~\cite{DBLP:journals/corr/abs-1111-4795} and PMIA~\cite{pmia} are strong among proxy-based methods. See a recent survey~\cite{Li2018InfluenceMO} for details.


There exist two on-going studies~\cite{DBLP:conf/atal/YanSLMLS19,DBLP:journals/corr/abs-1906-07378} most relevant to ours. 
In \cite{DBLP:conf/atal/YanSLMLS19}, a neural network-based influence function is used to predict whether a given node will activate or not. DISCO \cite{DBLP:journals/corr/abs-1906-07378} learns a function that estimates the influence of a given seed set and then selects seed nodes that maximize estimated influence based on deep reinforcement learning.
However, their learning models are {\it transductive}, i.e., incapable of estimating influence in social networks unseen during training. They can estimate influence only in a training network potentially with different activation probabilities. 
We aim at designing an {\it inductive} method, which is capable of estimating the influence of seed nodes in networks whose connections and activation probabilities are completely unseen during training.
In addition, our method estimates MC simulation results and thus can be equipped with greedy-based IM algorithms \cite{Kempe:2003:MSI:956750.956769,Goyal:2011:COG:1963192.1963217,6729575}, while the former~\cite{DBLP:journals/corr/abs-1906-07378} directly searches seed nodes. 

Machine learning approaches have been used for different but relevant  tasks, including to predict the future size of a cascade from its initial stage \cite{8264864} and to predict whether a node has been activated or not given its near neighbors' activation states and a subgraph around the node \cite{qiu2018deepinf}.
There also exist many attempts to predict solutions of various NP-hard problems, and our work was greatly inspired by them. Solutions for the satisfiability, maximal independent set, minimum vertex cover, traveling salesman, and knapsack problems can be predicted by deep learning models~\cite{Dai:2017:LCO:3295222.3295382,2019arXiv190303332M,NIPS2015_5866,DBLP:journals/corr/BelloPLNB16,DBLP:journals/corr/ZophL16}.

\section{Discussion on Linear Threshold}\label{sec:lt}
We briefly discuss why we focus on the IC model but exclude the LT model in this work. As described earlier, we use an upper bound information to ease the prediction task. For the LT cascade model, however, it is not easy to find tight upper bounds~\cite{6729575}. If $\mathbf{u}_{i}$ is not tight, our design not to directly predict the raw values will be discouraged. In our experience, directly predicting raw values is inferior to our design relying on an upper bound.  For this reason, we leave influence prediction under LT as an open problem.

\section{Conclusions}
\label{sec:conclusion}
In this work, we present MONSTOR, an inductive learning algorithm for estimating the influence of seed nodes under the IC model.
In our experiments, MONSTOR accurately estimated the influence even in networks unseen during training.
Moreover, simulation-based influence maximization algorithms equipped with MONSTOR, which replaces repeated MC simulations, performed reliably well, outperforming state-of-the-art competitors in 17 out of the 27 considered cases.

\section*{Acknowledgements}
This work was supported by National Research
Foundation of Korea (NRF) grant funded by the Korea government
(MSIT) (No. NRF-2020R1C1C1008296), the Yonsei University Research Fund of 2020-22-0074, and Institute of Information
\& Communications Technology Planning \& Evaluation (IITP) grant
funded by the Korea government (MSIT) (No. 2019-0-00075, Artificial Intelligence Graduate School Program (KAIST) and No. 2020-0-01361, Artificial Intelligence Graduate School Program (Yonsei University)).
This research was results of a study on the ``HPC Support'' Project, supported by the `Ministry of Science and ICT' and NIPA.

\bibliographystyle{IEEEtran}
\bibliography{IEEEabrv,sample-base}

\end{document}